\documentclass[aps,pr
,preprint
,showpacs
,groupedaddress
,superscriptaddress
,nobibnotes
]
{revtex4-1}
\usepackage{amsmath,amsthm,amssymb}
\usepackage{array}
\usepackage{subfigure}
\usepackage{graphicx}
\usepackage{fancyhdr}
\usepackage{xcolor}
\usepackage{extarrows}
\usepackage{enumerate}
\usepackage{epstopdf}
\usepackage{bm}
\usepackage[colorlinks,
          linkcolor=black,
            citecolor=black,
            urlcolor=blue
           ]{hyperref}
\usepackage{natbib}
\usepackage{comment}

\usepackage{color}

\usepackage{xr}
\usepackage{tabularx}

\begin{document}
\title{Hall anomaly by vacancies vs fragments of vortex lattice: Quantitative analyses of new evidences}
\author{Ruonan Guo}
\affiliation{National Laboratory of Solid State Microstructures, Nanjing University, Nanjing 210093, China}
\author{Yong-Cong Chen}
\email{chenyongcong@shu.edu.cn}
\affiliation{Shanghai Center for Quantitative Life Sciences \& Physics Department, Shanghai University, Shanghai 200444, China}
\author{Da Jiang}
\affiliation{Institute for Frontiers and Interdisciplinary Sciences,Zhejiang University of Technology, Zhejiang 310014, China}
\author{Ping Ao}
\email{aoping@sjtu.edu.cn}
\affiliation{College of Biomedical Engineering, Sichuan University, Chengdu, Sichuan Province, Sichuan Province 610044, China}
\begin{abstract}
Despite numerous recent studies on the Hall anomaly following the discovery of cuprate superconductivity, the origin of this phenomenon remains contentious. We demonstrate that a previously proposed mechanism, in which vacancy-on-fragment of the flux-line crystal, provides an alternative explanation for the observations of $\rm{Bi_{2}Sr_{2}CaCu_{2}O_{x}}$ thin films made by Nitzav and Kanigel [Phys. Rev. B. 107, 094516 (2023)], without the need for adjustable parameters. Specifically, we show that the power-law behavior of $\rho_{xy}$ over $\rho_{xx}$, with and without sign reversal, is consistent with the picture of vacancies versus fragments. Interestingly, we find that the effective length of vortex lines is consistently 1.5 unit cells (UC) across different experiments, independent of film thickness.

\end{abstract}

\maketitle

\renewcommand{\vec}[1]{\mathbf{#1}}
\newcommand{\bvec}[1]{\mbox{\boldmath $#1$}}
\newcommand{\cmmnt}[1]{\ignorespaces}

The Hall anomaly, which manifests in the mixed-state Hall resistivity of type-II superconductors, has been the subject of numerous theoretical studies. However, these studies fail to reach a consensus on the origin\cite{nozieres1966motion,hagen1990anomalous,hagen1993anomalous,zhu2001correlation,wang1991anomalous,hall1956rotationII,hall1956rotation,nozieres1966motion,hagen1990anomalous,
hagen1993anomalous,wang1991anomalous,ao1993berry,thouless1996transverse,xu2002scaling,shi4,josephson1965potential,richter2021resistivity,yu2019high} . While a lack of progress in experimental techniques may be partially responsible for this, a recent elegant experiment reported by Nitzav and Kanigel \cite{nitzav2023hall} stands out.
The high-temperature superconductivity significantly extends the parameter range where anomalous Hall effect can be observed, paving the way for improved analysis. They measured the longitudinal and Hall resistivity of an overdoped ($p=0.19$, $T_{c}=77$ K) pristine sample under magnetic fields ranging from 0.4 to 14 T, with and without sign reversal in the transverse resistivity as the temperature drops below $T_{c}$. Apparently, their data support the results of another recent experiment \cite{zhao2019sign} and the latter was examined based on a sophisticated many-body model, a vacancy-on-fragment mechanism involving the vortex lattice \cite{ao1998motion}. In a previous work of ours~\cite{guo2022hall}, excellent agreement was found between the model and experimental data, with no adjustable parameters.

The model attributes the Hall anomaly to the motion of vacancies in pinned and de-pinned fragments of the flux-line crystal. In this communication, we present an in-depth analysis of the activation energies and power law relationship deduced from the new report \cite{nitzav2023hall}. The experimental data are extracted and compared with the theoretical prediction of the energies of the conduction "carriers" in the superconducting BSCCO thin films, under varying magnetic field. The excellent agreement between them indicates that the report is supportive to the vacancy-on-fragment mechanism. In addition, the fitting suggests that the effective length for the vortices in BSCCO is consistently 4.5 nm or 1.5 UC (unit cell height), regardless of the actual sample thickness, a point that is worth pursuing in future studies.

Hall resistivity sign inversion occurs in between the Kosterlitz-Thouless transition temperature for the vortex crystal, $T_{KT}$, and the critical temperature, $T_{c}$, a regime where local fragments of vortex lattice with quasi long-range order preserved likely exist. As a consequence, the concept and presence of vacancies largely apply within a fragment, which may be pinned and depinned. To quantitatively compare the measured activation energy, $E_{a}$, from the slope of the curves in the Arrhenius formula with the theoretical one from vacancy formation on the flux-line crystal, we plot the logarithm of longitudinal resistance $\rho_{xx}$ versus $1/T$. The average values are tabulated in Table~\ref{Tab2}. As the thermal activation of dissipative vacancies in the superconductor film is the dominant source of resistance $R$ (i.e., $\rho_{xx}$), we have
 \vspace{-0.5cm}
\begin{multline}\label{logarithm}
 \;\;R=A\exp\left(E_a/k_{B}T\right) \;\Rightarrow\; \\
 \log_{10} R(T)=\log_{10} A-(\log_{10}e)[E_a(T)/k_BT],\;\;\;\;
\end{multline}
where the pre-factor $A$ and activation energy $E_{a}$, which are both included in the exponential term, may depend weakly on temperature $T$ in addition to the explicit Boltzmann factor $k_B T$. In \cite{guo2022hall}, $E_{a}$ was identified as the vacancy formation energy,
 \begin{eqnarray}
 \label{energy}
 E_v(T) &=& \frac{d\ln 2}{2\sqrt{3}\pi}\left(\frac{a}{\xi}\right)\left(\frac{\Phi_0}{4\pi\lambda(T)}\right)^2.
\end{eqnarray}
Moreover, in the Ginzburg-Landau (GL) theory, the penetration depth $\lambda(T)$ can be expressed as follows
\begin{equation}\label{lambda}
 \lambda(T)=\lambda(0)/[1-(T/T_c)]^{1/2}.
\end{equation}

\begin{figure}[htbp]
{
\includegraphics[width=1\textwidth]{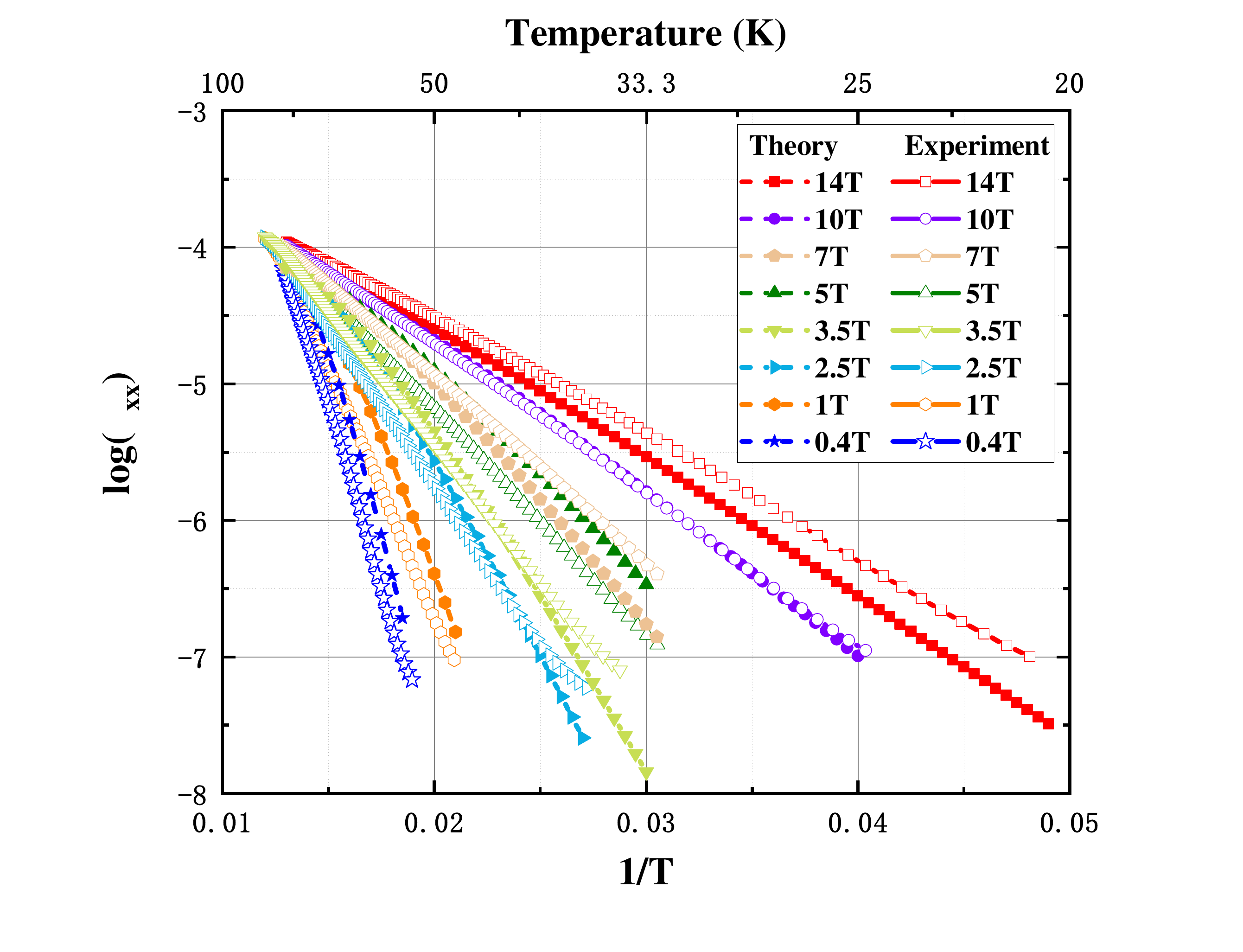}}
\protect\caption{An Arrhenius plot the resistance versus temperature, with solid lines and symbols representing experimental data from \cite{nitzav2023hall}, and open symbols and dashed lines representing data from Eq.(\ref{logarithm}). The various lines correspond to different external magnetic fields, controlled by vacancy excitations with the activation energy described in Eq.(\ref{energy}).
}
\label{Arrhenius2}
\end{figure}

To proceed further, we take from \cite{prozorov2000measurements} $\lambda(0)=2690$ \AA ~for BSCCO-2212, which may be compared to possible values from i) reversible magnetization, $\lambda\approx 2100$ \AA ~\cite{kogan1993role}; ii) uSR, $\lambda\approx 1800$ \AA ~\cite{lee1993evidence}; and iii) lower critical field measurements, $\lambda\approx 2700$ \AA ~\cite{niderost1998lower}. $T_c=77$K and $\log_{10} A\approx -4.15$ are read off from the experimental figure for the BSCCO-2212 film \cite{zhao2019sign}. Other experimental parameters include the GL parameter $\kappa=86$ \cite{stintzing1997ginzburg}, the coherence length $\xi=\lambda(T)/\kappa$, together with the flux quantum of Cooper pair $\Phi_0=hc/2\left|e\right|=2.07\times 10^{-7}$ G$\cdot$ cm$^2$ and $k_{B}=1.38\times 10^{-16}$ erg/K. The effective film thickness $d=1.5 $ UC $=2\times 1.5\times 15.35\times 10^{-8}$ cm (where 15.35 \r A  is the half height a unit cell in Bi-2212 ~\cite{dou2018experimental}), same as the effective thickness used in Ref.~\cite{guo2022hall}. All quantities are in CGS units.

We then apply Eq.~(\ref{logarithm}) to Fig. 2(a) in Ref.~\cite{nitzav2023hall}, with the results shown in Fig. \ref{Arrhenius2}. The theoretical values of the average energy of vacancy formation under diverse strength of magnetic field are presented in Table~\ref{Tab2}. Both the figure and the table display excellent agreement between theory and experiment, with no adjustable or extra fitting parameters.

\begin{table}[!ht]
\centering
\begin{tabular}{|c|c|c|c|c|c|c|c|c|}
  \hline
  B/T
  & 0.4 &	1 &	2.5 & 3.5	& 5 & 7 &	10 & 14 \\
  \hline
  Experimental $E_{a}$ value (K)
  & 510 &	356 &	230 &	196 &	164 & 135 &	108 & 87 \\
  \hline
  Theoretical $E_{a}$ value (K)
  & 479 &	344 &	255 &	225 &	142 & 160 &	109 & 97 \\
  \hline
 \end{tabular}
\protect\caption{The vacancy excitation energy of the BSCCO film was determined under various magnetic field. Experimental values were extracted using Eq.(\ref{logarithm}), while theoretical values were calculated based on Eq.(\ref{energy}).}
\label{Tab2}
\end{table}

With Eq.~(\ref{logarithm}), the slope of the curves in Fig. 1(b) in \cite{zhao2019sign} matches to $E_a/k_B$. The average activation energy across the temperature range is tabulated in Table~\ref{Tab2}. Our analysis indicates that vacancies are the primary contributor to the activation energy for $B\neq 0$, while the influence of independent vortices or vortex pairs is negligible (cf. \cite{guo2022hall}). The excellent agreement between the experimental and theoretical values for the longitudinal resistance across the full range of magnetic field suggests that the Hall anomaly results from many-body effects of vortex interaction involving vacancies on fragments of the flux-line crystal. Notably, the BSCCO-2212 bulk has a physical thickness of approximately 200 nm, but its effective thickness, as far as vortex dynamics are concerned, is only 4.5 nm and consistent with the value used in Ref.~\cite{guo2022hall}. Further experiments can be designed for better inspection.

We next explore the power-law behavior predicted by the vacancy-oriented model. In the low-temperature regime, both the longitudinal and the Hall resistivity decay exponentially with $T$ but the two are linked by\cite{ao1998motion}
 \begin{equation}\label{17}
 \rho_{xy} \propto \rho^\nu_{xx}.
\end{equation}
The vacancy origin of resistivity further predicted \cite{ao1998motion} that the exponent $\nu$ is given by
\begin{equation}\label{21}
 \nu = \frac{2a_v+b_v}{a_v+b_v}.
\end{equation}
The exponent $\nu$ is determined by the numerical factors $a_v$ and $b_v$ and can range between 1 and 2 depending on the specific details.
The range is supported by data from many studies \cite{shi1,shi2,shi3,shi4,shi5,nakai2011direct,PhysRevB.83.134506,wordenweber2017engineering,ghenim2004transport,xu2002scaling,mawatari1999anisotropic,yun1998anomalous,zechner2018transverse}.
If all vacancies are produced by pinning force, then $b_v=0$ and $\nu=2$.
Hence, if the value of $\nu$ is close to 2, the pinning origin will dominate the relation, which has crucial implications for the subsequent analysis.

 \begin{figure}[htbp]
{
\includegraphics[width=1\textwidth]{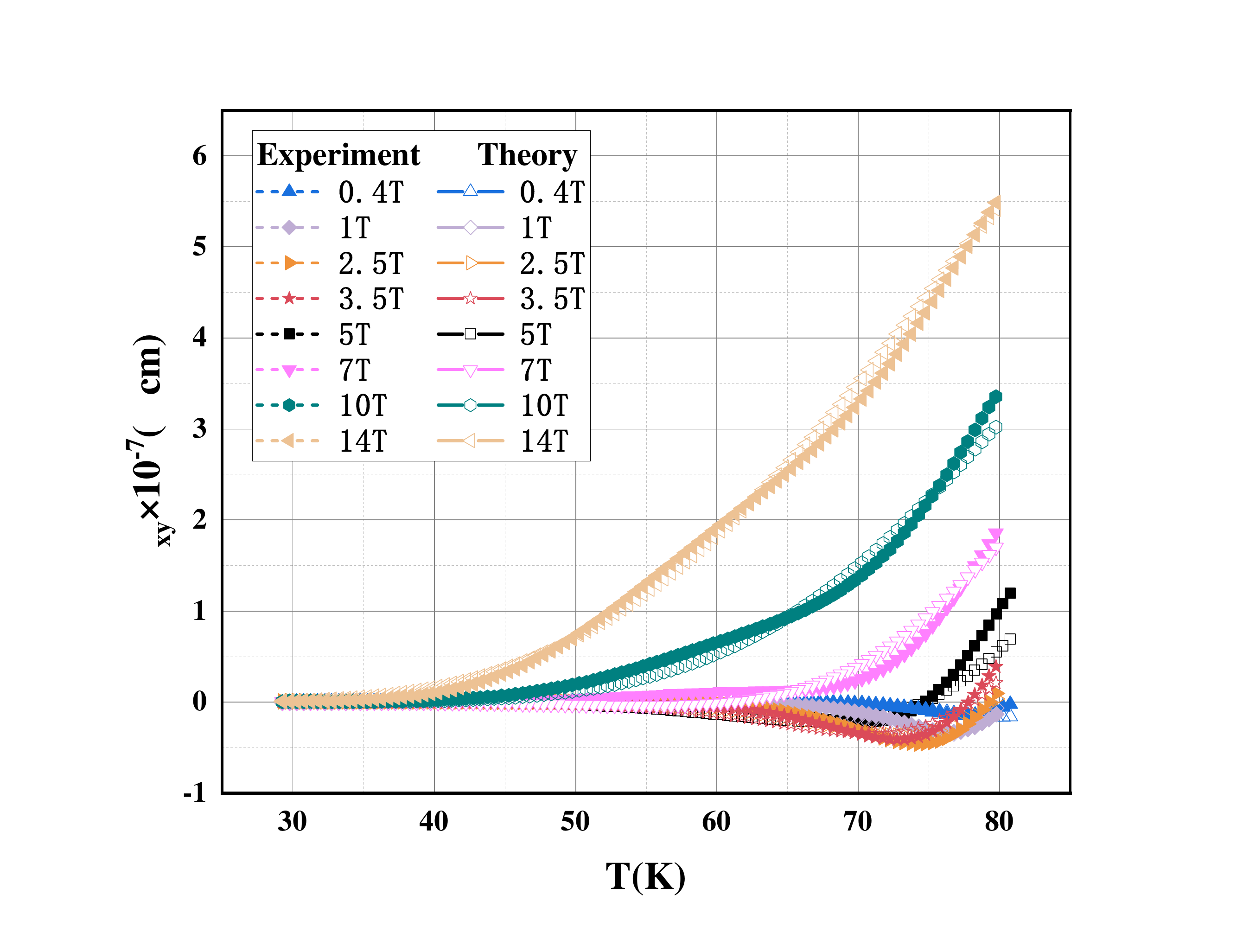}}
\protect\caption{ A plot of the Hall resistance versus temperature. The open symbols and dashed lines are from the experimental data reported in \cite{nitzav2023hall}, while the solid lines and symbols correspond to the results obtained using Eq.~(\ref{22}).
 }
 \label{v1+v2}
\end{figure}

The fragment mechanism applies in the temperature regime above the KT transition for the vortex crystal. The phase transition leads to fragments of vortex crystal floating around, consequently there are two competing types of "charge" carriers under electric field, fragments of crystal and the vacancies on the fragments. In terms of electrical property, a crystal fragment behaves the same sign as the normal carrier, which is positive in the current context. In contrast, a vacancy on a piece of crystal carries the opposite charge hence is negative. Both contribute to the Hall resistivity and the combined yields,  $\rho_{xy}$. Following Eq.~(\ref{17})
\begin{equation}\label{22}
 \rho_{xy} = G_{1}\rho^{\nu_{1}}_{xx}-G_{2}\rho^{\nu_{2}}_{xx}.
\end{equation}
The parameters $G_{1}$ and $v_{1}$ correspond to the fragments of vortex crystal, while $G_{2}$ and $v_{2}$ are associated with the vacancies on the fragments, as described in Eq.~(\ref{22}). We used the experimental values of $\rho_{xy}$ and $\rho_{xx}$ from Fig. 2(b) in Ref.~\cite{nitzav2023hall} to fit Eq.~(\ref{22}) with the least square errors on the steepest descent algorithm. The fitting results are displayed in Fig.~\ref{v1+v2}. Remarkably, despite the large variation in the magnitude of $\rho_{xx}$, the parameters $G_{1}$, $G_{2}$, $\nu_{1}$, and $\nu_{2}$ remain almost constant over a wide range of magnetic field, as shown in Table~\ref{Tab3}, with $\nu$'s very close to 1.8 (slightly different from the value obtained by the other group \cite{zhao2019sign}).
 \begin{table}[!ht]
\centering
\begin{tabular}{|c|c|c|c|c|c|c|c|c|}
  \hline
  B/T
  & 0.4 &	1 &	2.5 & 3.5	& 5 & 7 &	10 & 14 \\
  \hline
  $\nu_{1}$
  & 1.787 &	1.794 &	1.794 &	1.794 &	1.793 &    1.791 &	1.789 & 1.790 \\
  \hline
  $\nu_{2}$
  & 1.785 &	1.785 &	1.785 &	1.786 &	1.785 &	1.786 &	1.789 & 1.796 \\
  \hline
  $G_{1}$
  & 0.993 &	0.972 &	0.972 &	0.971 &	0.975 &	0.984 &	1.003  & 1.030 \\
   \hline
  $G_{2}$
  & 1.007 &	1.028 &	1.028 &	1.029 &	1.025 &	1.0157 &	0.997  & 0.971 \\
  \hline
 \end{tabular}
\protect\caption{The numerical values of exponents in a power-law relation from Eq.~(\ref{22}) on a least square fitting with steepest descent algorithm provided by MatLab\copyright.}
\label{Tab3}
\end{table}

The mechanism underlying the Hall anomaly is now quite transparent under the vacancy-on-fragment picture. The vacancy-related longitudinal resistivity is proportional to the number of thermally excited vacancies in the crystal fragments. Moreover, having $\nu$ close to 2 indicates that the excitation is mostly due to the pinning force on a fragment. The greater the number of vacancies produced, the less the pinning effects on the fragment itself, resulting in fragment-movement induced resistivity that is also proportional to the density of vacancies. Thus, the mobility of both carrier classes has a common origin, which naturally explains why they always have almost the same $\nu$ index and equal-magnitude charge densities in a wide range of parameter spaces, such as temperature, magnetic field, and resistivity. At high magnetic fields, the crystal constant becomes smaller, resulting in a higher density of vacancies. However, as the fragments become more crowded, the "hole" properties of the vacancies diminish, giving fragment movement a competitive advantage. Conversely, at lower densities of vacancies, the "holes" have an advantage when other pinning forces are present that can reduce the mobility of the fragments.

To summarize, our quantitative analysis, which does not involve adjustable parameters, sheds light on the Hall anomaly in the presence of vortex lattice and offers an alternative explanation to the experiment conducted by Yuval Nizav and Amit Kanigar \cite{nitzav2023hall}. Specifically, the power-law behavior in the low-temperature regime can be satisfactorily interpreted through the vacancy-on-fragment mechanism. We found a universal effective thickness of the vortex dynamics of 1.5 UC, which warrants further investigation.

One of the authors (RG) would like to acknowledge the helpful discussions on this research with Prof. Yuval Nitzav at Technion-Israel Institute of Technology. This work was partially supported by the National Natural Science Foundation of China under Grant No. 16Z103060007 (PA).

\end{document}